\documentclass[conference]{IEEEtran}
\IEEEoverridecommandlockouts

\usepackage{cite}
\usepackage{amsmath,amssymb,amsfonts}
\usepackage{algorithmic}
\usepackage{graphicx}
\usepackage{textcomp}
\usepackage{xcolor}

\usepackage{bbm}
\usepackage[utf8]{inputenc}
\usepackage[english]{babel}
 \usepackage{graphicx,epstopdf}
\usepackage{amsthm}
\usepackage{stfloats} 
\usepackage[ruled,vlined]{algorithm2e}
\usepackage{comment}

\def\BibTeX{{\rm B\kern-.05em{\sc i\kern-.025em b}\kern-.08em
    T\kern-.1667em\lower.7ex\hbox{E}\kern-.125emX}}
    
\begin{document}

\title{Low-Latency MAC Design for Pairwise Random Networks}

\author{Irshad~A.~Meer$^*$, Woong-Hee Lee$^+$, Mustafa~Ozger$^*$, Cicek Cavdar$^*$, and  Ki Won Sung$^*$\\
$^*$KTH Royal Institute of Technology, Sweden $^+$ Korea University, South Korea  \\
Email: \{iameer, ozger, cavdar, sungkw\}@kth.se, woongheelee@korea.ac.kr}

\maketitle

\begin{abstract}Feasibility of using unlicensed spectrum for ultra reliable low latency communications (URLLC) is still a question for beyond 5G wireless networks. Low latency access to the channel and efficiently sharing spectrum among the multiple users are the main requirements for exploiting unlicensed spectrum for URLLC. Listen before talk and back-off procedures implemented to avoid the collisions in channel access hinder the low latency communication. In this paper, we propose a novel low-latency medium access control (MAC) scheme based on the collision resolution for a pairwise random wireless network. We use geometric sequence decomposition for collision resolution among the competing users. This enables the system to tackle collisions and thus removing the need for carrier sensing and back-off procedures. This saves time in obtaining access to the channel and improves the efficiency of the system. We implement our approach in the synchronized time slotted system and show that it yields significant improvement over existing MAC schemes.
\end{abstract} 


\section{Introduction}
  
Industry 4.0 has provided increased digitalization and automation where robots or machines have to communicate with each other to work efficiently and in synchronization with each other \cite{oesterreich2016understanding}. One of the essential components of the 5G-and-beyond mobile networks is the connectivity in Industry 4.0 with strict requirements on the latency, availability, and reliability \cite{stefanovic2018industry}. The use of different technologies in Industry 4.0 such as autonomous machines, advanced robotics, edge computing  and machine learning based feedback loops, are being explored \cite{azari2018self,azari2019risk,masoudi2020device}. Industry 4.0 works with connected, data-sharing components and thus communication with low latency and higher reliability is of prime importance.  
In connected vehicles, machines, or robots, latency in the wireless access plays an important role for enabling the time critical operations. Time sensitive information has to be transmitted quickly and reliably. In order to fulfil these demands, it is clear that a “one size fits all architecture” like cellular 5G will not be sufficient. Also, given the saturation in the licensed bands because of the growing number of users (terrestrial, aerial, or industrial), fulfilling their requirements for latency and reliability will be a challenge \cite{nshimiyimana2016comprehensive}. This has precipitated the need to look and innovate in the unlicensed band for enabling the low latency communication. 

In this paper, we explore the use of unlicensed bands for ultra  reliable low latency communications (URLLC). Unlicensed bands with current medium access control (MAC) like carrier-sense multiple access with collision avoidance (CSMA-CA) in Wi-Fi are reliable but have high latency. Listen before talk (LBT) and back-off mechanisms that are employed in unlicensed bands hinder the use of URLLC services in these bands. Therefore, in order to use URLLC services in unlicensed bands, one of the main challenges is to design a new MAC scheme that enables low latency access in unlicensed bands.

There is much of the research documented to show the inefficiency in terms of time lost in accessing the channel for standard IEEE 802.11 and proposed ways to improve it \cite{lee2015making,lei2019differentiated}. Different approaches like collision avoidance or tuning the parameters such as back off time, transmission length and slot time  for optimization of CSMA are proposed in \cite{cali2000dynamic, siris2006optimal, kim2004throughput}. In \cite{lee2015making}, authors propose a new protocol based on the function of demand-supply differential of link capacity captured by the local queue length for making IEEE 802.11's distributed coordination function (DCF) near optimal. Among the efforts to improve the performance of the IEEE 802.11 DCF, we have two competing arguments to handle the collisions; 1) Given a collision, should the transmitter be more aggressive in trying to transmit the data \cite{rajagopalan2009network,liu2010towards,ni2011q} or 2) the transmitter goes into the exponential back-off and thus becoming less aggressive\cite{wu2002performance}. Authors in \cite{li2011crma} also investigate the problem of resolving collision by leveraging the random linear codes. However, the decoding algorithm is limited by the number of available channels and it needs to have a perfect channel estimation. 

Investigations into the access delay experienced by the vehicle in connecting to an on road Wi-Fi access point are considered in \cite{xu2017delay}. They concluded that  
the access delay increases almost linearly with the number of contending nodes. Authors in \cite{8643807} investigate how throughput is affected by the access procedure in a road side Wi-Fi network. All of these investigations have pointed out low latency is not guaranteed in a Wi-Fi like system. This is because of the inherent delay caused by the medium access which uses collision avoidance techniques. Therefore, to make unlicensed spectrum feasible for the URLLC, we need to modify the MAC scheme or develop a new scheme to reduce the average access delay.

In this paper, we investigate the use of unlicensed spectrum for communication between nodes (which can be machines or any communicating device in Industry 4.0, or in vehicular communication). Since the devices mostly work in pairwise, we confine the scope of our study to a pair-wise random network.
We propose a new MAC scheme called geometric sequence decomposition - multiple access (GSD-MA). We use GSD, which is a computationally efficient mathematical technique for decomposing the colliding signals \cite{lee2020geometric}. By employing GSD, we let the collisions happen for the access request and grant messages so that they can be resolved at the receiver. This is in contrast to the existing MAC schemes relying on collision avoidance.



The paper is organized as follows. Section \ref{sec3} provides the system model for the pairwise random network and define the performance metric for our work and also discuss about the benchmark scheme. In Section \ref{sec4}, we provide the algorithm for our proposed scheme. Section \ref{sec5} provides simulation setup and the performance results of our proposed multiple access scheme. Finally, Section \ref{conc} concludes our paper.

 \begin{figure*}[t!]
	\centerline{\includegraphics[width = 0.8 \textwidth
	]{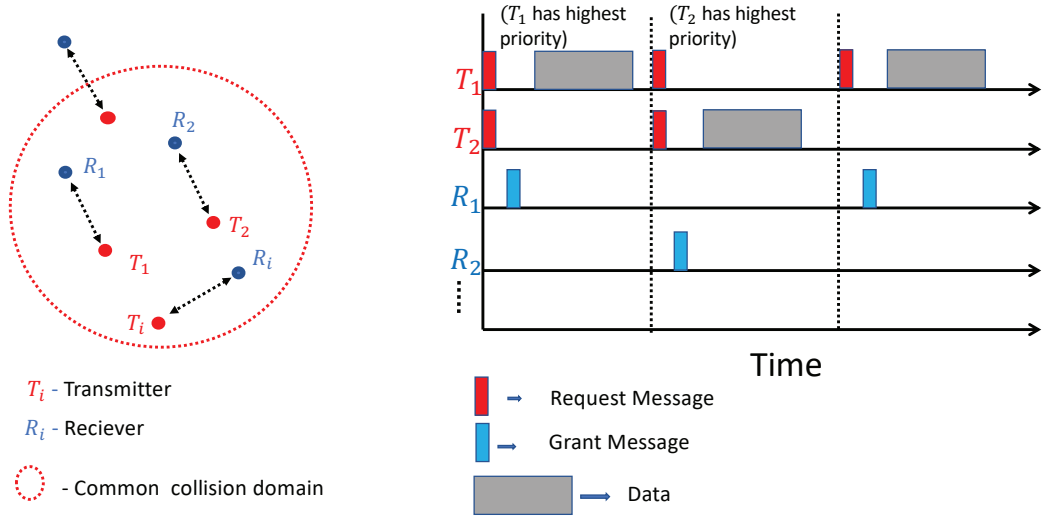}}
	\caption{ System model, time domain representation of GSD-MA.}
	\label{fig:F_1}
\end{figure*}

\section{System Model and Performance Metrics}
\label{sec3}
We consider a pairwise random network with fixed transmitters and receivers as shown in Fig. \ref{fig:F_1}. The transmitter and receiver in a pair are always within range of each other. The channel model considered is the free space path loss model. We consider a synchronous time-slotted system, where a  request message and a grant message are of single time slot duration. All the individual pairs are autonomous and are not coordinated by a centralized entity. This makes the MAC design follow a distributed approach for the random access. All the notation are provided in Table \ref{table:notation}.

To analyse the performance of our proposed scheme, we use two performance metrics, which are defined as follows:

\noindent \textbf{Efficiency:} We define efficiency, $E_f$, as the ratio of the number of slots a node is transmitting the data to the  total number of slots where the node is transmitting including the waiting and error handling time. If 
$n$ represents the number of slots a node is transmitting, $k$ as the number of slots a node is waiting for the others with higher priority to transmit and  $m$ as the number of slots the node encounters errors in transmitting the request or grant messages which depend on the symbol error rate. Then, $E_f$ will be given as:
\begin{equation}
    E_f = \frac{n}{(n+k+m)}.
    \label{efficiency}
\end{equation}
The nodes will not always be able to transmit data  because of the 1) priority of some other node in getting access to the channel, and 2) error in receiving the request/grant message. When an error occurs, the sender will repeat the transmission in the next slot. All these factors will lead to the loss of time (slots). While in case of the CSMA-CA, nodes will not be able to transmit because of the 1) listen before talk, 2) collisions, and 3) the back-off. 

\noindent \textbf{ Access Delay:} Access delay is the average time a node has to wait in order to start data transmission. Here, we ignore the transmission delay or the delay experienced in the other layers of the network. If $P_s$ denotes the success probability of the request/grant message in the network, $P_c$ is the probability that more than one transmitter is transmitting at the same time, $P_p^i$ is the probability that node $i$ has highest priority, and $T_s$ represents the slot time in our system. Then, the probability that node $i$ has access to the channel is given by:
\begin{equation}
    P_i = (1-P_c)P_s^2 + P_c P_p^i P_s^2.
    \label{P_delay}
\end{equation}
We have two scenarios for successful transmission, 1) when there are no multiple transmissions at the same time and there is no error in transmitting the request or grant messages, the probability to transmit is given by $(1-P_c)P_s^2$, 2) when there is a collision and node $i$ has highest priority, then the probability to transmit successfully is given by $P_c P_p^i P_s^2$. Let $n_i$ be the slot in which node $i$ gains access of the channel for the first time. The probability that $n_i = k$ is
\begin{equation}
    \mathrm{Pr}(n_i = k) = (1-P_i)^{k-1} P_i,
\end{equation}
where the probability $P_i$ is defined in \eqref{P_delay}. The average access delay for node $i$ is:
\begin{equation}
    D_i = \left(\mathbb{E}[n_i] - 1 \right)T_s,
    \label{delay}
\end{equation}
where $\mathbb{E}[.]$ is the expectation operator. 

\begin{table}[t!]
\caption{Key Notations Used.} 
\centering 
\resizebox{\columnwidth}{!}{
\begin{tabular}{ l   l  } 
\hline\hline 
\textbf{Notation} & \textbf{Description}  \\ [0.5ex] 
\hline 

$T_{s}$  & Time slot (sec) \\ 
$\lambda$   & Arrival rate (packet/sec) \\
$N_p$   & Number of node in the system  \\
$n$   &  Number of slots for which a node is transmitting \\
$k$   &  Number of slots for which a node is waiting \\
$m$   &  Number of slots in which an error occurred \\
$P_{c}$   & Probability that a request/grant message is successful   \\
$P_{s}$   & Probability that more than one transmitter \\ 
 &  is transmitting at the same time \\[1ex] 
\hline 
\end{tabular}
}
\label{table:notation} 
\end{table}

\section{Proposed MAC Scheme}
\label{sec4}
\subsection{Main Approach}
To design the proposed GSD-MA, we use a mathematical method for decomposition of non-orthogonal superposed geometric sequences, called geometric sequence decomposition with $k$- simplexes transform (GSD-ST) \cite{lee2020geometric}. In wireless communications, the equidistant samples of a radio wave comprise a geometric sequence. Thus, aggregation of multiple radio waves can be viewed as the superposition of multiple geometric sequences. GSD-ST turns the problem of decomposing $k$ geometric sequences into  solving of a $k$-th order polynomial equation. This is done by transforming an observed sequence (signal) to multiple $k$-simplexes in a virtual $k$-dimensional space and then correlating the volumes of the transformed simplexes. This means that GSD-ST is capable of demodulating the colliding radio signals. The performance of GSD depends on the signal-to-noise ratio (SNR). Therefore, it works better in the high SNR environment. 

With GSD-ST, a collision of radio waves from a few transmitters can be effectively resolved by a receiver, thus giving us the ability to utilize the spectrum more efficiently. In the proposed GSD-MA, GSD-ST which enables collision-free reception of uncoordinated signals is employed for short access request and grant messages. Then, conventional schemes like orthogonal frequency division multiplexing (OFDM) are used for data transmission as shown in Fig. \ref{fig:F_1}. 
The key design ideas of GSD-MA are summarized as follows:
\begin{itemize}
    \item We separate the channel access and the data transmission parts in time domain.
    \item In order to obtain the access to the channel, each transmitter will transmit a request message containing the transmitter's ID and the priority to the receivers. 
    \item Each receiver in a common collision domain where everyone can hear each other receive request messages from multiple transmitters and will decode the request messages by means of GSD-ST.
    \item A receiver will send back a grant message only if its counterpart transmitter has the highest priority. 
    \item A transmitter decodes all grant messages, and transmits the data packet if it has been granted with the highest priority. 
\end{itemize}

Since the multiple access request and grant messages can be sent simultaneously, our approach removes the need for LBT, congestion windows and back-off mechanism. The priority in the access request message is set based on two methods. The first method requires each node to send the quantized local queue-length information in the accesses request message. The second method is based on the pre-assigned priority to each node guaranteeing that there are no nodes having the same priority with another node. It does not require nodes to transmit queue-length information but send a predefined priority index \cite{gupta2009low}. The process of the GSD-MA is described in Algorithm~\ref{algo}.


An advantage of the proposed GSD-MA is that it addresses the hidden node and exposed node problems without any further extension as described in the following subsections.

\begin{algorithm}[t!]
 
\nl For request messages sent simultaneously;\\
\nl \For{ all the receivers in range}{
-Decodes all request messages;\\
\If{Decoding is successful}{
- \If{transmitter $i$ has highest priority}{
- send grant message for $i$ to all;
}
}
}
\nl \For{ all the transmitters in range}{
-Decodes all grant messages;\\
\If{Decoding is successful}{
- \If{transmitter $i$ has highest priority}{
- Transmit the data packet;
}
}
}
\caption{Process of Proposed Scheme }\label{algo}
\end{algorithm}

\subsection{Hidden Node Case:} Hidden node is a classical problem in the wireless networking where multiple data packets from different transmitters which are not in range of each other are transmitted to a single receiver, thus resulting in the collision at the receiver. In our scheme, we handle the hidden node problem without any further overhead messages. As shown in Fig. \ref{fig:F_3}(a), if we assume that $T_1$  has higher priority than $T_2$. $R_1$ will receive both requests from $T_1$  and $T_2$ and gives grant to $T_1$. $R_2$ receives request from $T_2$ only and thus gives grant to $T_2$. $T_1$ receives $R_1$ grant only thus transmits. $T_2$ receives both $R_1$ grant and $R_2$ grant and refrains from transmitting.

\subsection{Exposed Node Case:}
\begin{figure}[t!]
	\centerline{\includegraphics[width =0.5 \textwidth]{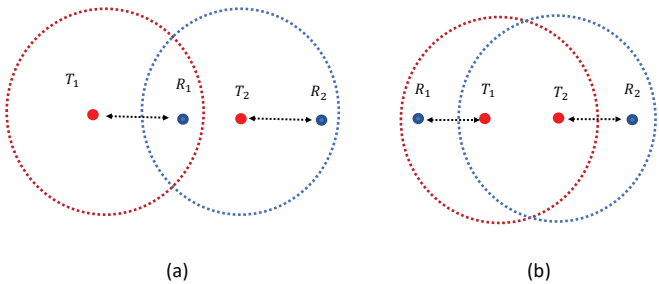}}
	\caption{(a) Hidden node and (b) Exposed node.}
	\label{fig:F_3}
	\vspace{-0.3cm}
\end{figure}

GSD-MA solves the exposed node problem as well. As shown in Fig. \ref{fig:F_3}(b), when $T_1$ and $T_2$ transmit the requests at the same time, they do not hear each other. $R_1$ receives the request from $T_1$ only and thus gives grant to $T_1$. $R_2$ receives the request from $T_2$ only and gives grant to $T_2$. Thus, both $T_1$ and $T_2$ receive grants and will transmit. 
\section{Performance Evaluation}
\label{sec5}
\subsection{Benchmark Scheme}
We compare the performance of our GSD-MA scheme with a benchmark, which in our case is CSMA-CA. In CSMA-CA, carrier sensing is used, and nodes start transmitting the data only after sensing the channel is idle. Once the transmission starts, the data packets are sent in their entirety. In this scheme, the node when it senses that the channel is being used, it will wait for a period of time (usually random) for the channel to be free before listening again for a free communication channel. Thus, the node loses time when it actually wants to send the data. Also, when the collision happens in the transmission, the node goes into a back-off procedure for a binary exponential random period before attempting to re-transmit. Thus, it again wastes the time when the node wants to transmit the data. Our scheme is different from the CSMA-CA as we do not have to wait for the channel to be free for obtaining the access. Based on the priority of the node, we can transmit our request to the target nodes. If a grant is given to a node, all the others will not transmit.
\subsection{Simulation Setup}
We consider a scenario that time is slotted and we have a common collision area with a number of node-pairs ranging from one to four. Our main focus is to compare the performance of our GSD-MA scheme with the benchmark, we use Matlab to implement the schemes. In our simulations, SNR ranges from $5$ to $60$ dB i.e., from a bad SNR region to good SNR region. The error probability in transmitting the request/grant messages is in the range of $0.001$ to $0.1$ in accordance with the results from \cite{lee2020geometric}. The error probability is the function of the SNR of the link and the number of node pairs transmitting at the same time. Since we are only interested in the access to the channel, we consider ideal error free transmission for the data. For the access request traffic, we consider a Poisson arrival traffic model for the nodes. We limit the number of node-pairs in our system model to four as the computational complexity in GSD increases for higher number of nodes \cite{lee2020geometric}.  We use legacy implementations of IEEE 802.11 with a slot time of $20$ micro-seconds, with a link speed of $10$ Mbps and packet size of $50$ slots. In case of the CSMA-CA implementation, we consider the Short Inter-frame Spacing (SIFS) of $8$ slots and DCF Inter-frame Spacing (DIFS) as $12$ slots.

\subsection{Simulation Results}
We provide the simulation results to show performance of our GSD-MA scheme by changing the number of nodes, data rates, and SNR values in the system. We also compare our scheme with the CSMA-CA to validate the performance. 
\subsubsection{Investigation of Efficiency with different number of nodes}
\begin{figure}[t!]
	\centerline{\includegraphics[width=0.47\textwidth]{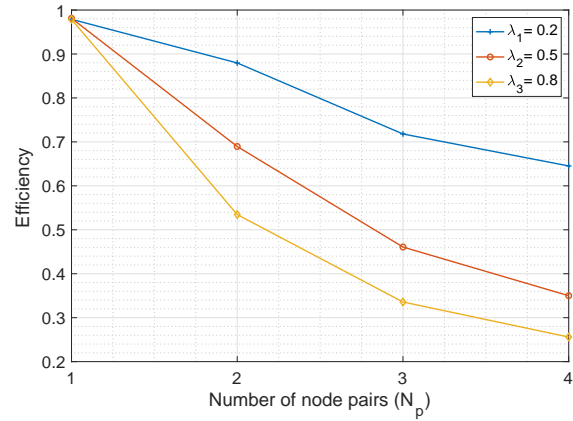}}
	\caption{Efficiency vs number of node pairs in the system for different arrival rates.}
	\label{fig:1}
	\vspace{-0.5cm}
\end{figure}

We investigate the effect of increased number of nodes in the system on the efficiency in Fig. \ref{fig:1}. An increase in the number of node-pairs from one to four results in a decrease in the efficiency. This is due to the two main reasons, 1)  As the number of nodes increase in the system, more time will be wasted in waiting for the channel to be free; 2) The error probability in transmitting/receiving the request and grant message is the function of number of simultaneously transmitting node pairs. As the node-pairs increase the error probability increases and time slots are wasted. The data packets arrive with a mean arrival rate of $\lambda$. We have defined three traffic loads, low traffic load $(\lambda_1= 0.2)$ means we have two new data packets per ten time slots. For moderate traffic load $(\lambda_1= 0.5)$, we have 5 data packets per ten time slots and for high  traffic load $(\lambda_1= 0.8)$, we have $8$ data packets per ten time slots. We observe that the traffic load affects the efficiency, the nodes have to wait more for others to transmit. A relatively low traffic load at four node-pairs in a system performs better than the system having only two node-pairs but with a higher traffic load.

\subsubsection{Investigation on Efficiency vs SNR}
\begin{figure}[t!]
	\centerline{\includegraphics[width=0.46\textwidth]{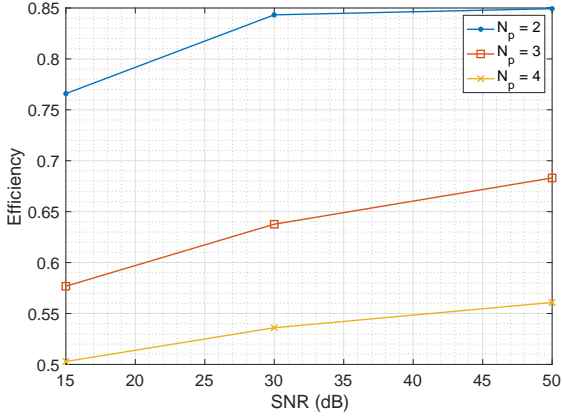}}
	\caption{Efficiency vs SNR with arrival rate of $\lambda_2= 0.5$.}
	\label{fig:2}
\end{figure}

The error probability in transmitting/receiving the request and grant message is the function of the SNR of the link. This can be seen from Fig. \ref{fig:2}, as the SNR increases i.e., the channel quality is better, the efficiency also increases. For different numbers of node-pairs in the system, we can see an increase in the efficiency with the increase in SNR. However the increase with three node-pairs $(N_{P}= 3)$ in the system is steeper as compared to two $(N_{P}= 2)$ and four node-pairs $(N_{P}= 4)$. This can be explained by the fact that with lower number of node-pairs, the channel efficiency is at its highest and can not be improved further. While with higher number of node-pairs, the main reason for loss of time slots is the channel being busy while other nodes are transmitting.  

\subsubsection{Investigation on average access delay for different arrival rates and SNR }
\begin{figure}[]
	\centerline{\includegraphics[width=0.46\textwidth]{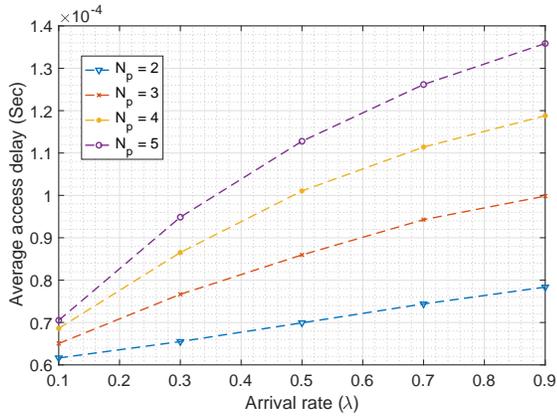}}
	\caption{Average access delay vs arrival rates for different number of node pairs.}
	\label{fig:10}
\end{figure}

\begin{figure}[t!]
	\centerline{\includegraphics[width=0.46\textwidth]{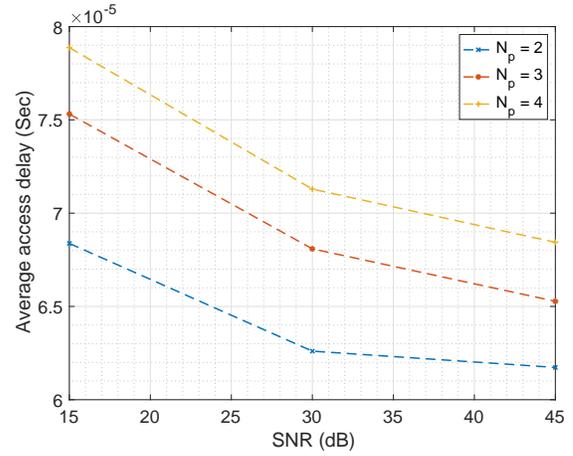}}
	\caption{Average access delay vs SNR for different number of node pairs.}
	\label{fig:11}
\end{figure}

We observe how arrival rates affect the average access delay in our proposed scheme for different number of nodes in Fig. \ref{fig:10}. As the arrival rates increases, average access delay increases. This can be explained by the fact that as more and more node-pairs want to send the data, they have to wait for the transmissions based on the priority. Thus, it leads to more and more average access delay. The average access delay is also function of the number of nodes in the system. As number of nodes increase, average access delay increases. Furthermore we can observe that the steepness of the curve, i.e., the rate of increase in delay in a system having more number of nodes is less as compared to the system having less nodes. This is because when the node pairs are more in number, an increase in arrival rate will not have much of an affect.  
We also investigated the average access delay as a function of the SNR of the link between the nodes in Fig. \ref{fig:11}. We can observe that delay variation for different SNR values is small as the error probability is small. The small error probability at the lower number of node-pairs does not reflect any change in the average access delay as the SNR is changed but at higher number of node-pairs, we can observe the change. 

\subsubsection{Comparison of GSD-MA with CSMA-CA}
\begin{figure}[t!]
	\centerline{\includegraphics[width=0.46\textwidth]{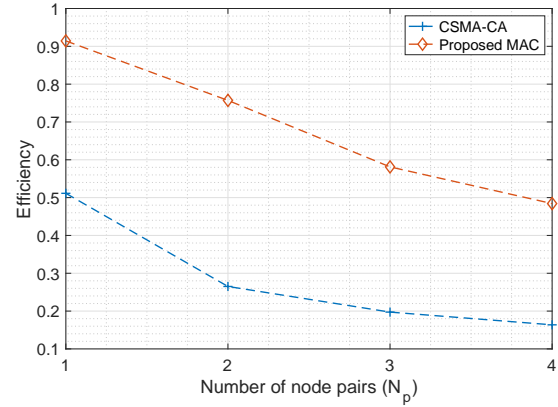}}
	\caption{Efficiency: CSMA-CA vs proposed MAC scheme with SNR = 30dB and $\lambda_2 = 0.5$.}
	\label{fig:3}
	\vspace{-0.3cm}
\end{figure}

In order to evaluate the performance of the proposed scheme we compare it with the benchmark scheme, i.e., CSMA-CA. As can be seen from the Fig. \ref{fig:3}, GSD-MA performs better than the CSMA-CA even at the higher number of simultaneously transmitting node-pairs and even much better at the lower number of node-pairs. Our scheme is 50\% more efficient than the CSMA-CA in terms of the time efficiency. Thus it is a perfect candidate for the low latency communication in the unlicensed bands. 

\begin{figure}[t!]
	\centerline{\includegraphics[width=0.46\textwidth]{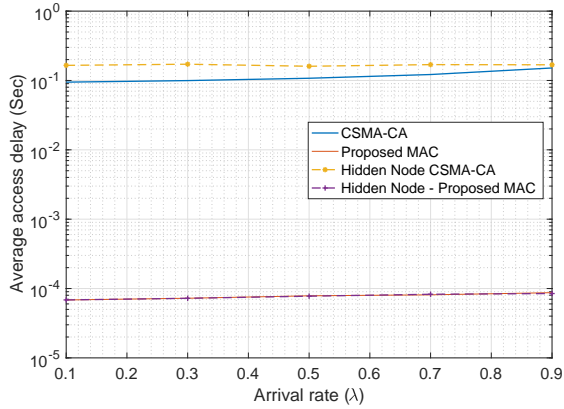}}
	\caption{ Delay Comparison of CSMA-CA vs proposed MAC scheme with SNR = 30 dB.}
	\label{fig:13}
	\vspace{-0.3cm}
\end{figure}

We also compare the performance in terms of the average access delay in Fig. \ref{fig:13}. We observe that the average access delay in the CSMA-CA case is much higher as compared to the proposed MAC. We also observe that hidden node problem has no effect on our scheme as both the curves overlap. In the case of CSMA-CA, hidden node problem causes around 15\% to 60\% increase in average access delay. The effect is more dominant when the arrival rate in small because at higher arrival rates the average access delay for CSMA-CA increases while remaining almost unchanged for the hidden node case. 
\section{Conclusion}
\label{conc}
In this paper, we study the feasibility of using unlicensed spectrum for ultra reliable low latency communications for beyond 5G wireless networks. We propose a novel low-latency medium access control scheme based on the collision resolution for a pairwise random wireless network called geometric sequence decomposition - multiple
access (GSD-MA). In this, we use mathematical technique of GSD for demodulating the non-orthogonally superposed radio waves. We remove the need for implementing collision avoidance techniques like listen before talk and back-off procedure and enable the collision resolution with the help of GSD. With our multiple access scheme, we can obtain access to the wireless channel faster. Also, it improves efficiency of the system. Simulation results demonstrate how the efficiency changes with respect to the number of nodes in the system and  SNR. Furthermore, we study the average access delay for different arrival rates, changing SNR, and different number of nodes in the system. For the comparison of the proposed GSD-MA with the conventional CSMA-CA, we analyze the efficiency and the access delay for the two schemes and show that our scheme outperforms CSMA-CA. Extension of the work for different network topologies and wireless environments remains as a further study.

\section*{Acknowledgment}
	
Results incorporated in this paper received funding from the ECSEL Joint Undertaking (JU) under grant agreement No 876124. The JU receives support from the EU Horizon 2020 research and innovation programme and Vinnova in Sweden.

\bibliographystyle{IEEEtran}
\bibliography{Ref}

\end{document}